\documentclass[sigconf,natbib=false,screen,]{acmart} 
\AtBeginDocument{%
  }

\setcopyright{acmlicensed}

\copyrightyear{2026}
\acmYear{2026}
\acmDOI{XXXXXXX.XXXXXXX}
\acmConference[SAC'26]{The 41st ACM/SIGAPP Symposium on Applied Computing}{March 23--27, 2026}{Thessaloniki, Greece}
\acmISBN{979-X-XXXX-XXXX-X/26/03}
\setcopyright{none}
\renewcommand\footnotetextcopyrightpermission[1]{} 
\usepackage[caption=false,font=footnotesize]{subfig}
\RequirePackage[
  datamodel=acmdatamodel,
  style=acmnumeric,
  ]{biblatex}

\begin{document}

\title{Towards Automated and Predictive Network-Level Energy Profiling in Reconfigurable IoT Systems}

\author{Mohammud J. Bocus}
\affiliation{%
  \institution{University of Bristol }
  \city{Bristol}
  \country{UK}
}
\email{junaid.bocus@bristol.ac.uk}

\author{Senhui Qiu}
\affiliation{%
  \institution{University of Bristol }
  \city{Bristol}
  \country{UK}
}
\email{senhui.qiu@bristol.ac.uk}

\author{Robert J. Piechocki}
\affiliation{%
  \institution{University of Bristol }
  \city{Bristol}
  \country{UK}
}
\email{r.j.piechocki@bristol.ac.uk}

\author{Kerstin Eder}
\affiliation{%
  \institution{University of Bristol }
  \city{Bristol}
  \country{UK}
}
\email{kerstin.eder@bristol.ac.uk}

\renewcommand{\shortauthors}{M.J. Bocus et al.}

\begin{abstract}
Energy efficiency has emerged as a defining constraint in the evolution of sustainable Internet of Things (IoT) networks. This work 
moves beyond simulation-based or device-centric studies to deliver measurement-driven, network-level smart energy analysis. The proposed system enables end-to-end visibility of energy flows across distributed IoT infrastructures, uniting Bluetooth Low Energy (BLE) and Visible Light Communication (VLC) modes with environmental sensing and E-ink display subsystems under a unified profiling and prediction platform.
Through automated, time-synchronized instrumentation, the framework
captures fine-grained energy dynamics across both node and gateway layers. 
We developed a suite of tools that generate energy datasets for IoT ecosystems, addressing the scarcity of such data and enabling AI-based predictive and adaptive energy optimization.
Validated within a network-level IoT testbed, the approach demonstrates robust performance
under real operating conditions. 
\end{abstract}



\maketitle
\pagestyle{plain}

\section{Introduction}
The Internet of Things (IoT) is a technology used in various
applications, such as industrial control, smart metering, home
automation, agriculture, and eHealth, where IoT devices need
to operate for extended periods under energy constraints, and it
is crucial for the applications to be aware of their own energy
consumption \cite{Hussain2017EnergyCO}. 
Therefore, the accurate measurement of energy consumption is widely recognized as a cornerstone for designing sustainable IoT networks, particularly as device deployments scale and application requirements become increasingly diverse.

In \cite{Katz_2024}, the SUPERIOT initiative was presented as a step toward building sustainable IoT ecosystems. The project introduced a holistic framework that brings together radio-frequency and optical communication to enable dual-mode connectivity, energy harvesting, and positioning capabilities. In parallel, \cite{lirias4159920} demonstrated the design of batteryless sensor platforms leveraging Bluetooth Low Energy (BLE) and light-based IoT (LIoT) technologies, powered through indoor photovoltaic harvesters. 
Beyond connectivity and energy autonomy, the SUPERIOT project also places emphasis on sustainability by investigating the use of printed electronics and conductive inks to reduce the environmental footprint of IoT devices across their lifecycle. 
Our contributions complement these efforts by delivering practical methodologies for energy-aware optimization and predictive analytics at the network-level in the SUPERIOT framework, thus tackling one of the most pressing challenges in IoT—efficient and reliable energy management. 
The main contributions of this work are:
\begin{enumerate}

 \item 
We extend our ongoing research on energy measurement, analysis and predictive modeling in reconfigurable IoT (RIoT) nodes, expanding the scope to include custom-engineered gateway and access point platforms for comprehensive network-level analysis.

\item We implement 
tools to accurately forecast the energy consumption of both RIoT nodes and gateways as well as access points across various operating states, scenarios, and configuration parameters. Beyond prediction, these tools can also generate realistic energy datasets based on user-defined scenarios, enabling the development and training of Artificial Intelligence (AI)/Machine Learning (ML)-based energy prediction models, enabling proactive energy management and optimization across distributed IoT networks. 

    \item 
We design and implement a fully automated energy profiling system that eliminates manual intervention and ensures consistent data acquisition across heterogeneous IoT nodes and network conditions.

\item 
Unlike prior methods focused solely on individual devices, our framework provides synchronized energy metrics at both micro (node-specific) and macro (network-wide) levels, capturing real-time interactions among devices, gateways, and communication modes (e.g., BLE, VLC).

\item 
We evaluate the proposed framework using real-world experiments with actual IoT nodes and gateways, demonstrating its scalability, reliability, and potential to guide sustainable IoT design.

\end{enumerate}

\section{Related Works}
The authors of \cite{PowerofModels} developed an analytical model for estimating the power consumption of wireless sensor nodes in IoT applications. 
The proposed model 
accounts for all energy expenditures, including communications, data sensing (acquisition), and processing. 
Their approach allows for a new framework to analyze energy life-cycles in applications, enabling engineers to understand the impact of different parameters on power consumption and make informed decisions about system design. 
The authors of \cite{park2003unified} introduced a novel simulation framework designed to assess power consumption comprehensively, encompassing both node-level performance and network-wide estimation. At the node level, the framework integrates a power simulator for the StrongARM processor with radio and sensing components. This integrated setup is incorporated into SensorSim \cite{SensorSim}, an enhanced version of the network simulator ns-2, to simulate realistic sensor network scenarios. The framework's capabilities are demonstrated through exploration of various power management schemes and interactions across node-level network layers and sensor field events.
The research in \cite{Gray2018EnergyCO} aims to understand and characterize the energy consumption of IoT services, particularly focusing on home automation, security, and video surveillance applications. Through empirical measurements and modeling, the author assesses the energy consumption of various components within IoT systems,
shedding light on their substantial energy demands compared to traditional household consumption. Additionally, the study investigates the energy efficiency of different wireless communication protocols commonly used in IoT applications and evaluates the potential of edge computing architectures to mitigate energy consumption challenges. 
In \cite{OZKAYA2024105009} the authors propose a fully simulated, model-based approach to estimate energy consumption at the system level, without the need for complete design, implementation, or physical measurements. 

Compared with prior studies, this work provides a distinct contribution through its empirical, measurement-driven analysis of custom-engineered, multi-functional RIoT nodes operating within a network. 
Existing studies on IoT energy consumption often rely on simulations or focus on isolated aspects such as sensing, computation, or specific communication modes. Such approaches fail to capture the complex, interdependent behaviors that emerge in multi-protocol, real-world deployments. 
For example, the existing approaches  often overlook broader macro-level influences—including network topology, the number of active nodes, communication protocols, and aggregate payload sizes—which can significantly affect overall energy demand~\cite{cristofani2024age}.
Conversely, many network-level measurement frameworks focus on macro-level parameters and tend to exclude detailed node-level characteristics`\cite{hammoud2025toward}. This separation of micro- and macro-level perspectives limits the ability to gain a comprehensive and precise understanding of energy consumption.

In contrast, this work presents a unified, network-level framework for automated energy measurement, analysis, and prediction in heterogeneous IoT systems. 
The proposed system enables 
energy profiling under realistic operating conditions. It also integrates an automated infrastructure that ensures consistent, synchronized data acquisition from multiple IoT nodes across diverse operating states. Beyond device-level analysis, the framework provides network-wide visibility by capturing multi-layer energy metrics encompassing BLE and VLC communication modes, gateway activity, and node interactions. Furthermore, it includes energy estimation and prediction tools capable of generating time-aligned, ML-ready datasets that support the training of models for accurate energy forecasting. The framework’s scalability, reliability, and applicability are demonstrated through extensive hardware-in-the-loop validation within our testbed, underscoring its potential for sustainable and intelligent IoT design.
\begin{figure}[t]
\centering
\subfloat[]{\includegraphics[width=0.40\linewidth]{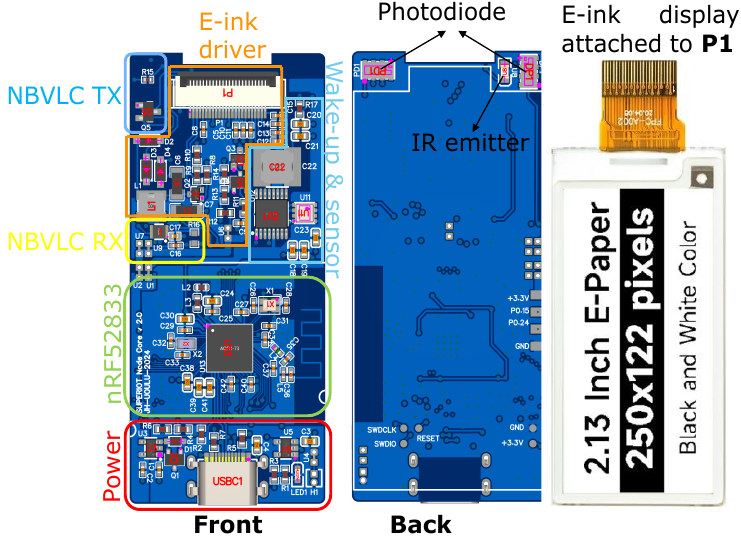}\label{fig:node}}
\hfil
\subfloat[]{\includegraphics[width=0.40\linewidth]{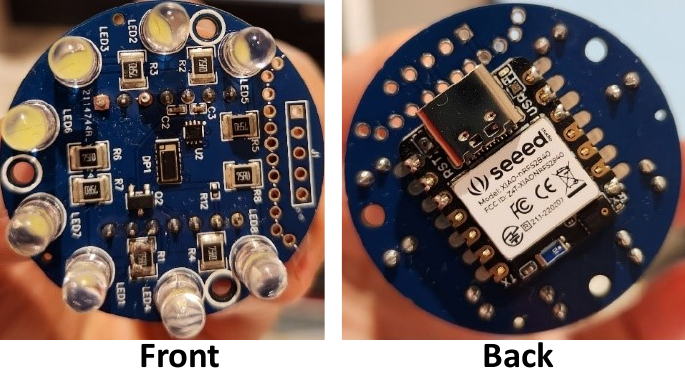}\label{fig:minilamp}}
\hfil
\subfloat[]{\includegraphics[width=0.35\linewidth]{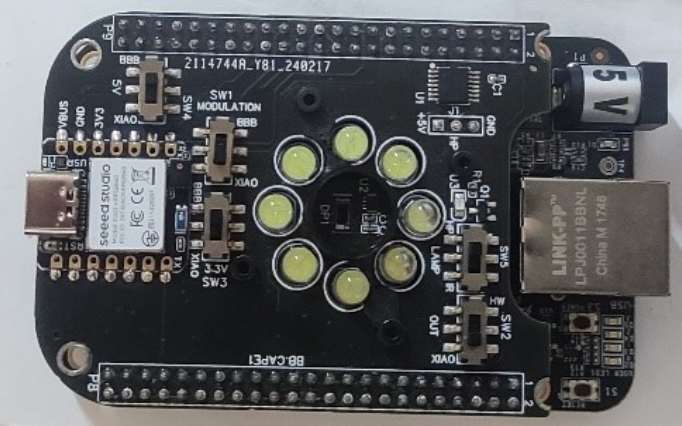}\label{fig:bbbap}}
\caption{Illustration of (a) Custom-engineered, multi-functional Si-based RIoT node, (b) Mini-lamp gateway supporting dual-mode communication (BLE and VLC), and (c) BBB Access Point (BBB platform + mini-lamp gateway mounted on a Cape).}
\end{figure}

\begin{figure}
    \centering    \includegraphics[width=1\linewidth]{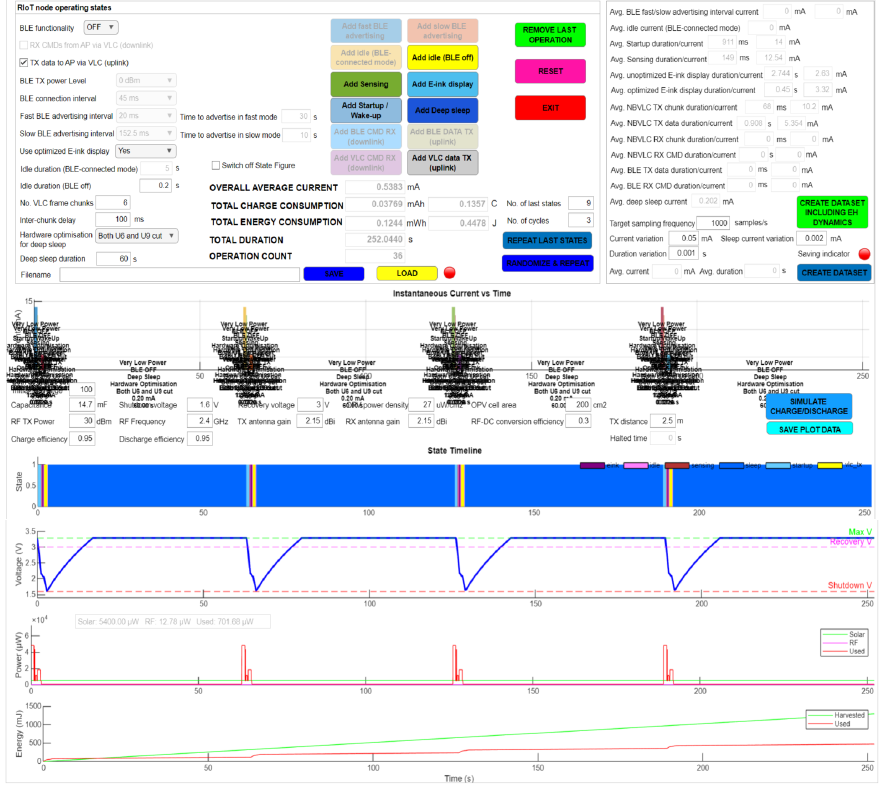}
    \caption{Node-level energy consumption prediction App. 
    }
    \label{fig:nodeapp}
\end{figure}
\section{System Description}
\subsection{SUPERIOT Hardware}
    Figure \ref{fig:node} shows the prototype silicon (Si)-based reconfigurable IoT node developed as part of the SUPERIOT project. 
At its core is the nRF52833 Bluetooth Low Energy (BLE) System on a Chip (SoC), which serves as both the main controller and wireless communication interface. For visible light communication (VLC), the node features a VLC transceiver, consisting of an infrared (IR) Light-Emitting Diode (LED) transmitter and a receiver built from an IR sensor preamplifier coupled with a photodiode. Environmental monitoring is enabled by the BME688 sensor, which accurately measures temperature, humidity, pressure, and gas levels, while the AS3933-BTST component provides a low-power light-based wake-up and timer function that enhances energy efficiency. A 2.13-inch black-and-white E-ink display (250 × 122 pixels) provides an energy-efficient visual output.
   
    The mini-lamp gateway developed in the SUPERIOT project (see Figure \ref{fig:minilamp}) is designed to interface with the RIoT node in Figure  \ref{fig:node}, facilitating the exchange of data and commands via both VLC and BLE. The mini-lamp gateway consists of a BLE-enabled Seeed Arduino nRF52840 SoC and photodiodes and LEDs for providing VLC functionality. 

    The main access point developed as part of the project consists of a BeagleBone Black (BBB) platform with the custom-engineered mini-lamp gateway (see Figure \ref{fig:minilamp}) integrated on a BBB Cape, as shown in Figure \ref{fig:bbbap}. The access point communicates using BLE and VLC protocols with the RIoT nodes. 
    This platform can be connected to a broader backbone network comprising routers, switches, and both locally hosted (e.g., Raspberry Pi-based) and cloud-based MQTT brokers, enabling remote bidirectional exchange of commands and data within the IoT infrastructure.

\subsection{Node-Level Energy Measurement, Analysis, Modeling and Prediction}
In~\cite{bocus2025energyawareriotsystemanalysis}, extensive energy measurements were taken on the Si-based RIoT node, covering various operating states including idle, sleep, BLE communication, VLC, environmental sensing, and E-ink display updates.
It was demonstrated that the initial energy consumed by the Si-based node can be reduced by over 60\% through software-level optimizations when the node performs all its functionalities.
One critical improvement involved refining the E-ink display's driving waveform, which significantly decreased the energy required during display updates. Additionally, it was observed that the node can be configured into a very low-power mode when only sensing and E-ink display operations are active for short intervals before the system enters a deep sleep state for an extended period (while communication modules remain deactivated). Hardware configuration optimizations were also evaluated in this very low-power mode, and further reductions in energy consumption were achieved during the deep sleep state.
Measurement-based energy models were also developed for the RIoT node to predict energy consumption under different operating modes.
These models were thoroughly validated using new measurement data obtained from the node under various configurations and scenarios. 
Remarkably, the models achieved an accuracy exceeding 97\%, indicating their reliability for predicting energy usage in diverse node configurations.

Additionally, we extended the SUPERIOT vision by integrating the node's energy models into a practical tool (see Figure \ref{fig:nodeapp}) that predicts energy consumption based on user-configurable parameters.
The developed application 
allows users to define custom IoT node operation profiles for a wide range of scenarios, specifying operational states such as BLE, VLC, sensing, E-ink display, wake-up, idle, and sleep. These scenarios can be configured as either periodic duty cycles or randomly triggered events, with easily adjustable parameters for rapid assessment of their impact on energy consumption. The application also includes an energy harvesting simulation module supporting both radio-frequency and light sources, incorporating user-defined parameters and realistic supercapacitor charge/discharge dynamics linked to node operations. 
The application can also generate realistic RIoT energy datasets that reflect defined operation modes and simulated harvesting effects—datasets that can further be used to train AI-based models for on-node energy prediction and optimization.
The predictive capability of the tool will be crucial in designing robust nodes that can adapt their operation based on the energy available, thereby ensuring continuous and sustainable functioning in diverse deployment environments.
\begin{figure}
    \centering    \includegraphics[width=0.85\linewidth]{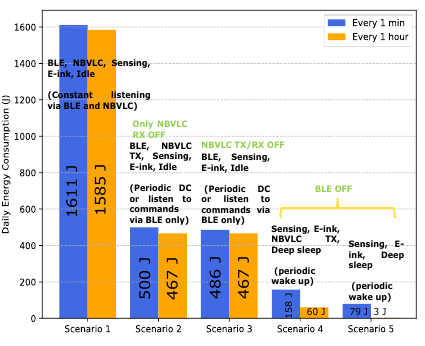}
    \caption{Predicted energy consumption of Si-based node over a 24-hour period under various scenarios. 
    }
    \label{fig:nodebar}
\end{figure}

\subsubsection{Energy Consumption Overview of RIoT Node}
We estimate the energy consumption of the RIoT node under different scenarios using the developed application illustrated in Figure \ref{fig:nodeapp}.

    \textbf{Scenario~1: BLE and VLC activated and the node continuously listens to the gateway commands.}
     In this scenario, the node performs all its core operations, including BLE, VLC, sensing, and E-ink display updates and receives commands (via both BLE and VLC) from the gateway. 
     The operational sequence for this scenario is as follows:
     The node powers on and initiates fast BLE advertising with a default interval of 20~ms. After 10~seconds, a gateway establishes a BLE connection with the node, at a BLE connection interval of 45~ms.
     The node remains in an idle BLE-connected state for another 5~seconds, maintaining a BLE TX power of 0~dBm. The gateway sends a BLE command to the node for the latter to initialize its BME environmental sensor. The initialization process takes approximately 916~ms. After a 10-second idle period in BLE-connected mode following startup, the gateway starts sending commands via VLC at regular intervals (e.g., every 1~minute or 1~hour). Basically, for each command transmitted by the gateway, the node will perform the following operations: Sensing (516~ms), idle (1~second), optimized E-ink display update (435~ms), idle (1~second), VLC transmission of data (908~ms). Then the node will be idle until the next command is received from the gateway after either 1~minute or 1~hour.
    As shown in Figure~\ref{fig:nodebar}, Scenario~1 represents the worst case in terms of power consumption. Specifically, considering a 24-hour period, the node consumes approximately 1611~J when receiving commands at 1-minute intervals and 1585~J when the interval is 1~hour. This elevated energy usage is primarily due to the timer functions responsible for Pulse Width Modulation (PWM) and encoding VLC data~\cite{bocus2025energyawareriotsystemanalysis}. 

    \begin{table}[t]
  \caption{Measured current and power of BBB Access Point under different conditions.}
  \label{tab:bbb_power_tests}
  \centering
  \footnotesize 
  \setlength{\tabcolsep}{2.5pt} 
  \begin{tabular}{c l r r}
    \toprule
    \textbf{ID} & \textbf{Condition} & \textbf{I (mA)} & \textbf{P (W)} \\
    \midrule
    1 & Boot: USB+Eth (VLC/BLE OFF) & 405 & 2.03 \\
    2 & Idle: USB+Eth (VLC/BLE OFF) & 255 & 1.28 \\
    3 & Idle: No USB/Eth (VLC/BLE OFF) & 170 & 0.85 \\
    4 & Idle: Eth only (VLC/BLE OFF) & 241 & 1.21 \\
    5 & TX: USB+Eth (VLC/BLE OFF) & 388 & 1.94 \\
    6 & TX: USB+Eth (VLC 98\%, BLE OFF) & 590 & 2.95 \\
    \bottomrule
  \end{tabular}
\end{table}
     \textbf{Scenario~2: BLE and VLC both on, and node performs tasks in a periodic duty cycle (but no constant listening to commands from the gateway).}
     This scenario closely resembles Scenario~1, with the key difference being that the node operates independently, following a periodic duty cycle for tasks such as sensing, E-ink display updates, and VLC data transmission to the gateway in the uplink, without listening for commands from the gateway.
 
     Alternatively, the node can also receive commands from the gateway (in downlink) via BLE only. Therefore, if the node does not need to listen for VLC frames from the gateway, the NBVLC functions (more details in~\cite{bocus2025energyawareriotsystemanalysis}) in the firmware can remain disabled and be activated only just before transmitting each VLC frame chunk. This approach leads to a significant reduction in energy consumption as shown in Figure~\ref{fig:nodebar}, with values around 500~J for a 1-minute duty cycle and 467~J for a 1-hour duty cycle over a 24-hour period. This represents approximately a 70\% reduction in energy usage compared to Scenario~1, while performing the same operations.

   \textbf{Scenario~3: BLE on and VLC off, and node performs tasks in a periodic duty cycle.}
In this scenario, the VLC functions are completely disabled for the whole operating duration, and all data transmissions are performed exclusively via BLE. The operational sequence and timing configuration remain consistent with Scenarios~1 and~2, preserving identical BLE parameters and duty-cycle intervals. 
As in the previous scenarios, the node initializes in BLE advertising and connected idle states, followed by BME sensor activation and a subsequent post-startup idle phase. As shown in Figure~\ref{fig:nodebar}, this scenario leads to energy consumption of approximately 486~J for a 1-minute duty cycle and 467~J for a 1-hour duty cycle over a 24-hour period. 
    
   \textbf{Scenario~4: BLE off and VLC on, and node performs tasks in a periodic duty cycle.} 
    In this scenario, the node performs its basic operations such as sensing and optimized E-ink display update, while BLE is deactivated and only VLC is used to transmit data in the uplink to a gateway in a periodic duty cycle (using the same VLC transmission strategy as in Scenario~2). After carrying out its main operations (active states), the node will go into a deep sleep state. 
    The operation duty cycle consists of the following sequence of states: Startup (911~ms), idle for 30~ms, sensing (149~ms), idle for 250~ms, optimized E-ink displaying (0.450~s), idle for 250~ms, transmission of a VLC frame (908~ms), deep sleep (1~min or 1~hour). Note that compared to the previous scenarios, the BME startup state will be part of the operation cycle and will be repeated over the node’s operating time. 
    By disabling BLE, relying solely on VLC for uplink data transmission, and allowing the node to enter deep sleep mode, Figure~\ref{fig:nodebar} shows a substantial reduction in energy consumption. Specifically, the node consumes approximately 158~J for a 1-minute duty cycle and 59~J for a 1-hour duty cycle over a 24-hour period. 
    
\textbf{Scenario~5: BLE off and VLC off, and node performs tasks in a periodic duty cycle.} 
Finally, in Scenario~5, both communication modules are deactivated. This means that there will be no transmission of data to a gateway, and the node performs only sensing and actuation tasks like E-ink display updates. Thus, the operation duty cycle consists of the following states: Startup (909~ms), idle for 30~ms, sensing (149~ms), idle for 250~ms, optimized E-ink displaying (0.544~s), deep sleep (1~min or 1~hour). 
As shown in Figure~\ref{fig:nodebar}, this approach achieves the lowest energy consumption, with values around 78~J for a 1-minute duty cycle and just 3~J for a 1-hour duty cycle over a 24-hour period. This scenario highlights how disabling communication modules and leveraging deep sleep states can dramatically reduce energy usage, making them ideal for applications where minimal power consumption is required.

\subsection{Gateway/Access Point Energy Measurement, Analysis, Modeling and Prediction}
\subsubsection{BBB AP Energy Consumption}
Table~\ref{tab:bbb_power_tests} summarizes the average current and power consumption of the BBB-based access point (AP) shown in Figure~\ref{fig:bbbap} under various operating configurations.
For test IDs 1–5, the VLC LEDs on the BBB CAPE were disabled, and BLE functionality on the Seeed Xiao nRF52840 chipset remained off.
During the initial boot sequence (test ID~1), the AP consumes approximately 405~mA (2.03~W) on average. In this configuration, a USB connection links the BBB’s USB~2.0 port to the Type-C interface of the Seeed Xiao nRF52840 module on the lamp CAPE, while an Ethernet cable connects the BBB AP to a network switch. The boot process lasts roughly 72~seconds, after which the system transitions to an idle state. In this state (test ID~2), with both USB and Ethernet connected, the current draw decreases to 255~mA (1.28~W).
Additional idle measurements were performed with only the Ethernet cable connected (test ID~4) and with both USB and Ethernet disconnected (test ID~3). Comparison of these results reveals that connecting the USB link between the BBB platform and the Xiao nRF52840 module increases the current by approximately 14~mA (tests~2 vs.~4), while adding the Ethernet connection raises it by roughly 71~mA (tests~3 vs.~4), without active Ethernet data transfer.
For active communication testing, iPerf3 was used to establish bidirectional Ethernet traffic between the BBB AP (client) and a laptop (server) connected through the same switch. Under continuous data transfer (test ID~5), the current consumption rises to 388~mA, representing an increase of about 133~mA relative to the idle condition in test ID ~2.
Finally, when the VLC LEDs operate at 98\%~brightness, the maximum power consumption recorded on the BBB AP (test ID~6) reaches approximately 2.95~W.

\begin{figure}
    \centering    \includegraphics[width=0.75\linewidth]{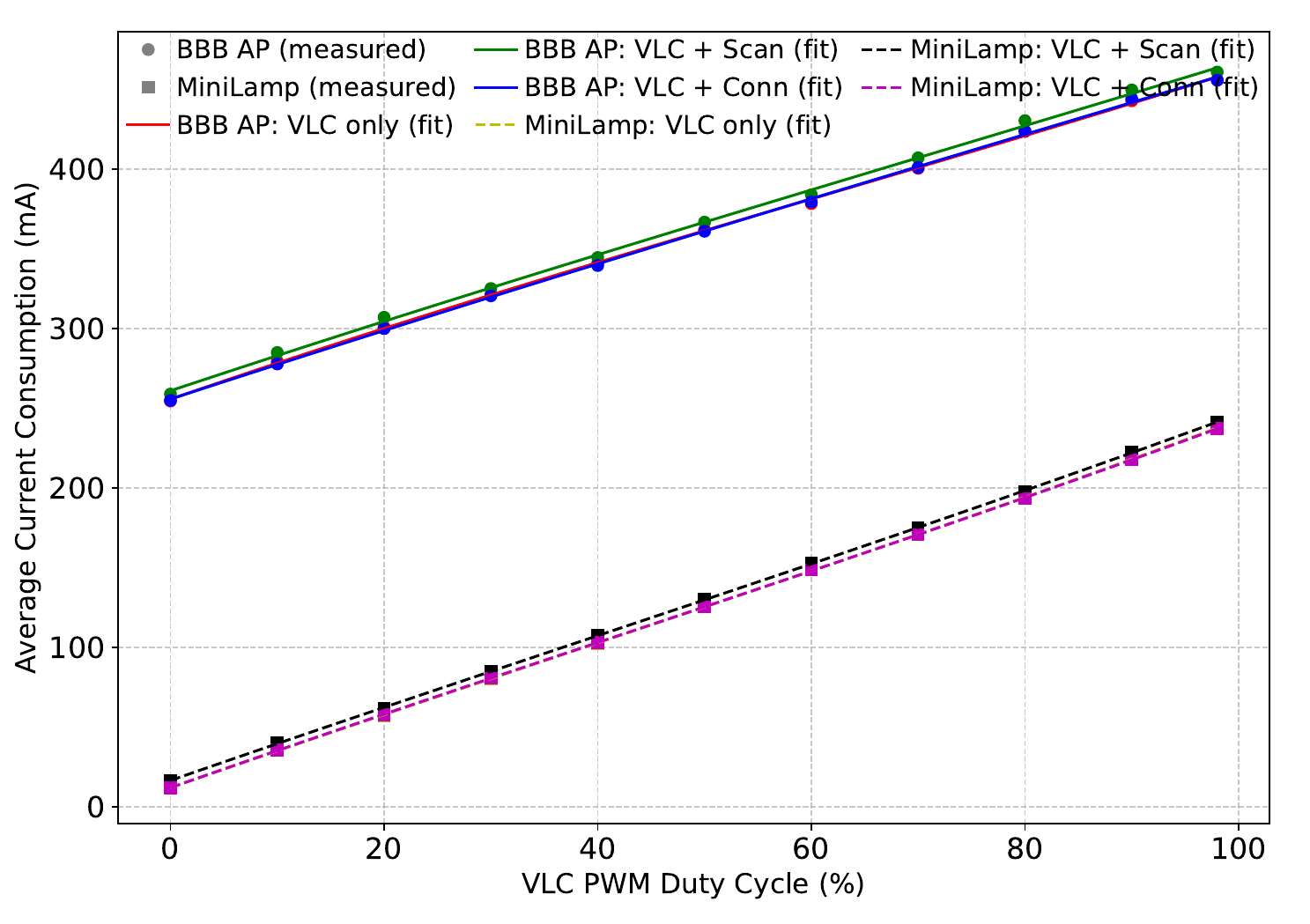}
    \caption{Comparison of the average current consumption between BBB AP and mini lamp gateway under three test cases. 
    }
    \label{fig:bbb_vs_mini}
\end{figure}

\begin{figure}[t]
\centering
\subfloat[]{\includegraphics[width=0.47\linewidth]{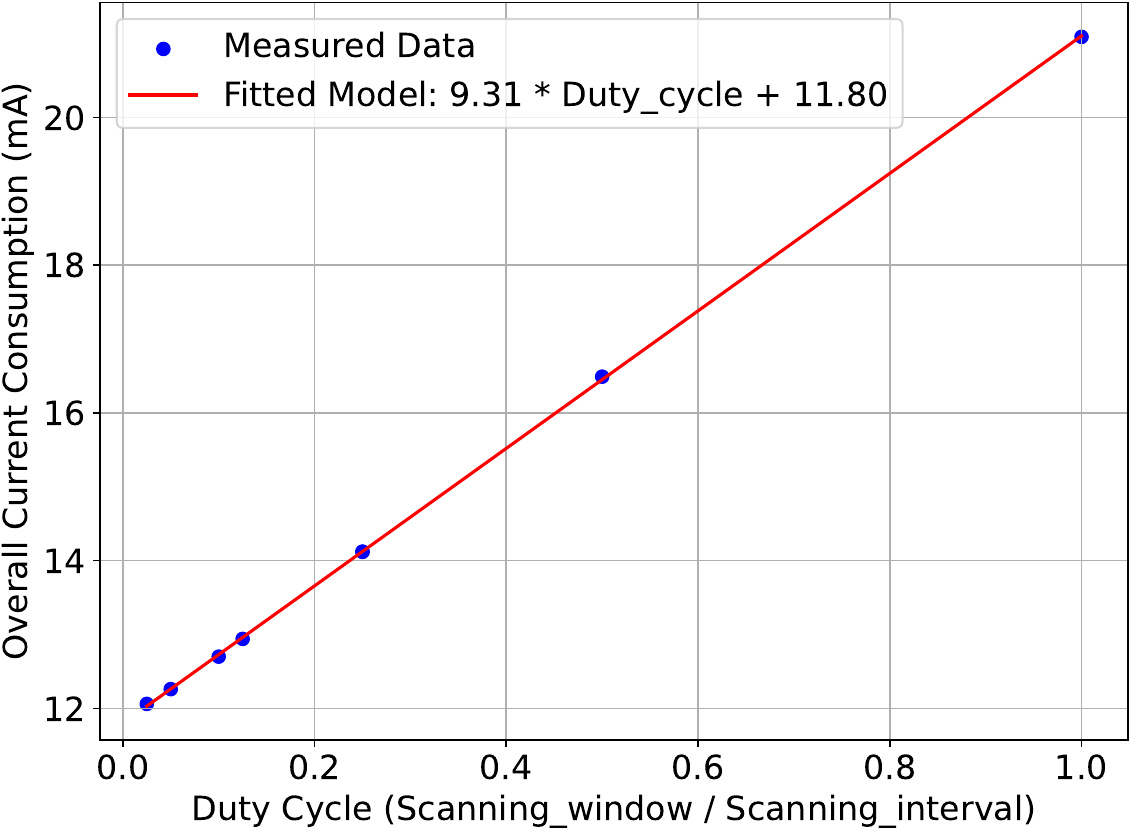}\label{fig:blescan}}
\hfil
\subfloat[]{\includegraphics[width=0.5\linewidth]{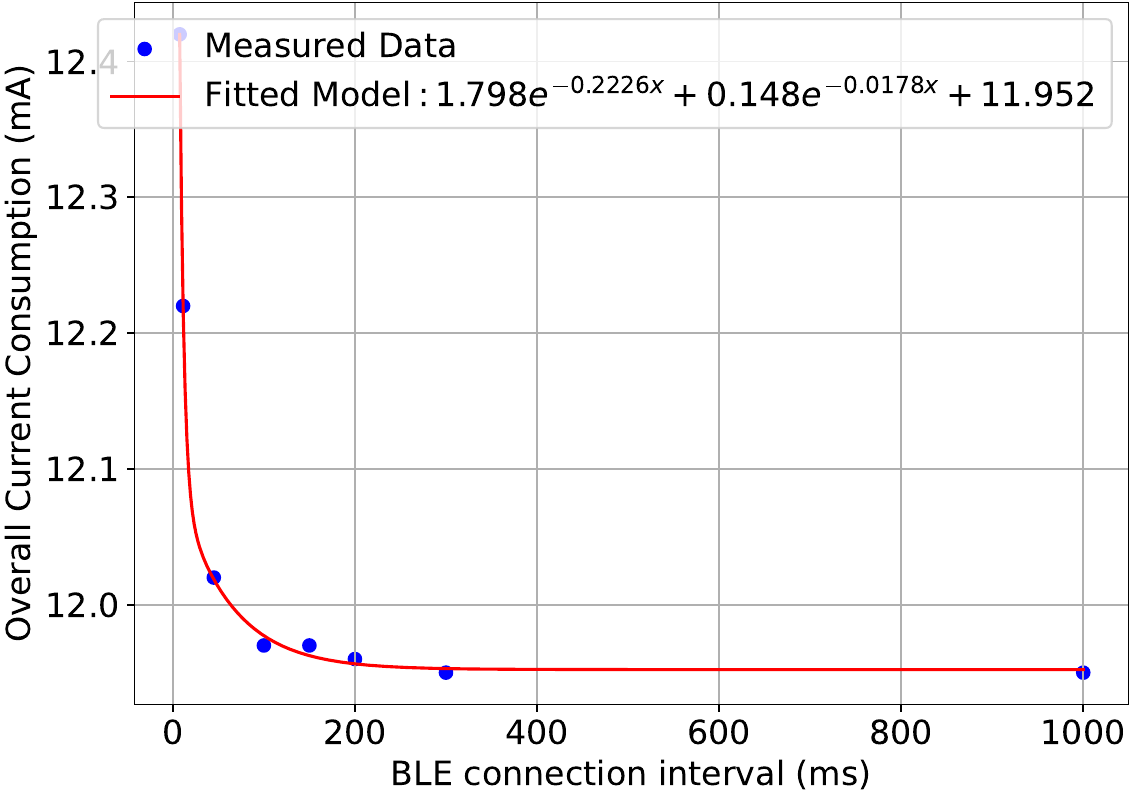}\label{fig:bleconn}}
\caption{Current consumption analysis of mini lamp gateway during different: (a) BLE scanning duty cycle values, (b) BLE connection intervals (VLC off in both cases).}
\end{figure}
\subsubsection{Impact of LED Brightness on Average Current Consumption}
To investigate the impact of illumination intensity on system power behavior, we varied the VLC PWM duty cycle— which controls the brightness of the lamp LEDs— and analyzed its effect on the average current consumption of both the BBB access point (AP) and the mini-lamp gateway. Figure~\ref{fig:bbb_vs_mini} compares the average current measured under three operating conditions:
(1) VLC enabled with BLE disabled;
(2) VLC active with BLE scanning (100~ms scan interval, 50~ms scan window); and
(3) VLC active with an established BLE connection to the RIoT node (connection interval = 45~ms).
Across all three test cases, the system remained in idle mode, with no data exchange between the gateway/access point and the node. The BLE transmit power was fixed at 0~dBm, and for the BBB AP, both the USB and Ethernet interfaces were connected but inactive (no Ethernet data traffic).
Both platforms exhibit a similar trend: the average current increases approximately linearly with the VLC PWM duty cycle, as shown in Figure~\ref{fig:bbb_vs_mini}. 

When BLE scanning is enabled alongside VLC, the current consumption experiences a nearly uniform upward offset relative to the VLC-only condition— approximately +5.31~mA for the BBB AP and +4.43~mA for the mini-lamp gateway. In contrast, when a BLE connection is established (connection interval = 45~ms) without active data transmission, the current difference relative to the VLC-only case remains negligible.
In Figure~\ref{fig:blescan}, we analyze the average current consumption of the mini-lamp gateway with the VLC LEDs turned off, under active BLE scanning on the chipset. Specifically, we investigate the impact of the BLE scanning duty cycle on the gateway’s energy consumption (BLE TX power level = 0~dBm). The results in Figure~\ref{fig:blescan} demonstrate a linear relationship between the scanning duty cycle and average current draw (with the fitted model closely following the measured data). As the duty cycle increases from 2.5\% to 100\% (continuous scanning), the mini-lamp gateway exhibits a corresponding increase in current consumption of approximately 75\%.
We further examine the average current consumption of the mini-lamp gateway when it is connected via BLE to a RIoT node, with no BLE data transfer and a TX power level of 0~dBm. As shown in Figure~\ref{fig:bleconn}, increasing the BLE connection interval from 11.25~ms to 1000~ms results in a slight decrease in average current consumption, from 12.22~mA to 11.95~mA. This variation is minimal, indicating that the current can be considered effectively constant over the tested range of connection intervals.

Overall, the BBB AP consistently consumes more current than the mini-lamp gateway, with an average excess of approximately +235~mA ($\approx$ +1.18~W) across all test conditions. This difference primarily reflects the higher baseline power requirements of the BBB platform compared to the lightweight mini-lamp gateway design.

\subsubsection{Energy Modeling}
Based on the measurements in Figure~\ref{fig:bbb_vs_mini}, we can derive a cubic polynomial equation for estimating the average current consumption of the BBB AP as a function of the VLC PWM duty cycle as follows (BLE off):
\begin{equation}
I_{\text{BBB\_vlc\_idle\_only}} \,(\text{mA}) = 3 \times 10^{-5} x^3 - 5.582 \times 10^{-3} x^2 + 2.319 x + 255.654 ,
\label{eq:vlc_only}
\end{equation}  
where $x$ represents the VLC PWM duty cycle (\%), in the range 0$\leq$ $x$$\leq$ 98\%. 
From Figure~\ref{fig:bbb_vs_mini}, we observe that the average current consumption of the BBB AP in the idle VLC mode and the BLE-connected mode (no BLE data transfer) is very similar to that of the VLC case only, i.e., 
\begin{equation}
I_{\text{BBB\_vlc\_idle\_and\_ble\_conn\_idle}} \approx I_{\text{BBB\_vlc\_idle\_only}} 
\end{equation}  
When both VLC and BLE scanning are active, we notice an almost parallel upward shift in current consumption (average slope in Figure~\ref{fig:bbb_vs_mini} is approximately 2.3~mA/\%)
when compared to the average current consumption for VLC only. 
Thus, Equation~\ref{eq:vlc_only} can be re-written as follows:
\begin{equation}
I_{\text{BBB\_vlc\_idle\_and\_ble\_scan}} \,(\text{mA})  = I_{\text{BBB\_vlc\_idle\_only}} + C ,
\label{eq:vlc_blescan}
\end{equation}  
where $C$ represents the average offset. %

Since the BBB AP exhibits a similar slope ($\sim 2.3~\mathrm{mA/\%}$) to the mini-lamp gateway when BLE scanning is active, we can approximate the BLE-scanning-only current equation for the BBB AP by adopting the mini-lamp gateway's slope during BLE scanning only ($9.30~\mathrm{mA/\%}$) from Figure~\ref{fig:blescan} and adjusting the intercept to account for the BBB AP's higher baseline consumption.
The average excess current consumption of the BBB AP relative to the mini-lamp gateway for VLC only (baseline) is:
\[
I_{\text{excess, VLC only}} = 254.54 - 11.97 = 242.57~\mathrm{mA},
\]
and for VLC + BLE scanning at a 50\% BLE duty cycle:
\[
I_{\text{excess, VLC + BLE scan}} = 258.91 - 16.49 = 242.42~\mathrm{mA},
\]
(values extracted from Figure~\ref{fig:blescan} and Figure \ref{fig:bbb_vs_mini} considering a VLC PWM duty cycle of 0\%).
Since the excess current remains approximately constant ($\sim 242.5~\mathrm{mA}$), we assume the BBB AP and mini-lamp gateway share the same slope, with the BBB AP simply having a higher base consumption due to the BBB platform itself (see Table~\ref{tab:bbb_power_tests}). Accordingly, the average current consumption of the BBB AP during BLE-scanning-only operation (VLC lamp off) can be approximated as:
\begin{align}
I_{\text{BBB\_ble\_scan\_only}} \,(\text{mA})  &= 9.3 \times BLE \ Duty \ Cycle + (11.8 + 242.5) \nonumber \\ &= 9.3 \times BLE \ Duty \ Cycle + 254.3,
\label{eq:blescanonly}
\end{align} 
where $BLE \ Duty \ Cycle = \frac{\text{BLE \ scanning \ window \ (ms)}}{\text{BLE \ scanning \ interval \ (ms)}}$.
Thus, Equation~\ref{eq:vlc_blescan} can be re-written as:
\begin{align}
I_{\text{BBB\_vlc\_idle\_and\_ble\_scan}} \,(\text{mA})  &= 3 \times 10^{-5} x^3 - 5.582 \times 10^{-3} x^2 + \nonumber \\ & 2.319 x +  254.3 + 9.3 \times BLE \ Duty \ Cycle.
\label{eq:vlc_blescan2}
\end{align}  
To evaluate the VLC transmission behavior, a frame composed of six 32-bit chunks (each lasting approximately 68 ms) was transmitted using the BBB access point, with BLE functionality disabled. The experiment was conducted across VLC PWM duty cycles ranging from 10\% to 90\%. The relationship between the VLC idle current and the average VLC transmission current (computed over the six frame chunks) is expressed as follows:
\begin{equation}
I_{\text{BBB\_vlc\_TX}} \,(\text{mA})
=
1.003 \times I_{\text{BBB\_vlc\_idle\_only}} + 0.4656 .
\label{eq:vlc_tx_idle}
\end{equation}  
Equation (\ref{eq:vlc_tx_idle}) indicates a near-linear relationship between the VLC transmission current and the VLC idle current. The slope being approximately unity implies that the VLC TX current is only marginally higher than the idle current, signifying minimal additional power overhead during transmission. 
We also derive the current (in mA) relationship between \textbf{VLC idle + BLE scanning} and \textbf{VLC transmission + BLE scanning}, based on measurements obtained across VLC PWM duty cycles ranging from 10\% to 90\%, as follows:
\begin{equation}
I_{\text{BBB\_vlc\_TX\_and\_ble\_scan}} \,(\text{mA})
=
0.9995 \times I_{\text{BBB\_vlc\_idle\_and\_ble\_scan}} + 3.113 .
\label{eq:vlc_tx_ble_scan}
\end{equation}  
The slope in Equation~\ref{eq:vlc_tx_ble_scan} is close to~1, meaning VLC transmission only marginally increases the power draw. 
The average current consumption during VLC data reception on the BBB AP, without and with BLE scanning, can be respectively approximated as follows:
\begin{equation}
I_{\text{BBB\_vlc\_RX}} \,(\text{mA})
\approx
I_{\text{BBB\_vlc\_idle\_only}}.
\end{equation}  
\begin{equation}
I_{\text{BBB\_vlc\_RX\_and\_ble\_scan}} \,(\text{mA})
\approx
I_{\text{BBB\_vlc\_idle\_and\_ble\_scan}}.
\end{equation}  
Next, we evaluate the energy consumption of the system when the BBB AP transmits commands to the node, and the node responds by sending the requested data (e.g., environmental BME sensor readings) back to the BBB AP via BLE. 
To isolate the BLE contribution, we vary the VLC PWM duty cycle (i.e., LED brightness) while keeping VLC communication inactive (VLC idle, no data transmission or reception). During this experiment, the BBB AP maintains a BLE connection with the node using a connection interval of 45~ms, and the AP’s default transmission power is set to 4~dBm. When a command is transmitted over BLE, the current profile exhibits a brief peak lasting approximately 4.5~ms. Similarly, receiving data from the node produces a shorter peak of about 2.5~ms. We measure the average current consumption associated with these peaks for both BLE command transmission and data reception at the BBB AP, across different VLC PWM duty cycle values ranging from 0\% to 98\% (VLC idle).
We model the relationship between the measured current (in~mA) for the \textbf{VLC idle + BLE-connected (idle)} and \textbf{VLC idle + BLE-connected (data reception)} states as follows:
\begin{align}
I_{\text{BBB\_vlc\_idle\_and\_ble\_RX}} \,(\text{mA})
&=
0.9915 \times I_{\text{BBB\_vlc\_idle\_and\_ble\_conn\_idle}} \nonumber \\ & + 106.1.
\end{align}
Similarly, we model the relationship between the measured current (in~mA) for the \textbf{VLC idle + BLE-connected (idle)} and \textbf{VLC idle + BLE-connected (command transmission)} states as follows:
\begin{align}
I_{\text{BBB\_vlc\_idle\_and\_ble\_TX}} \,(\text{mA})
&=
1.015 \times I_{\text{BBB\_vlc\_idle\_and\_ble\_conn\_idle}} \nonumber \\ & + 74.25.
\end{align}

\begin{figure}
    \centering    \includegraphics[width=1\linewidth]{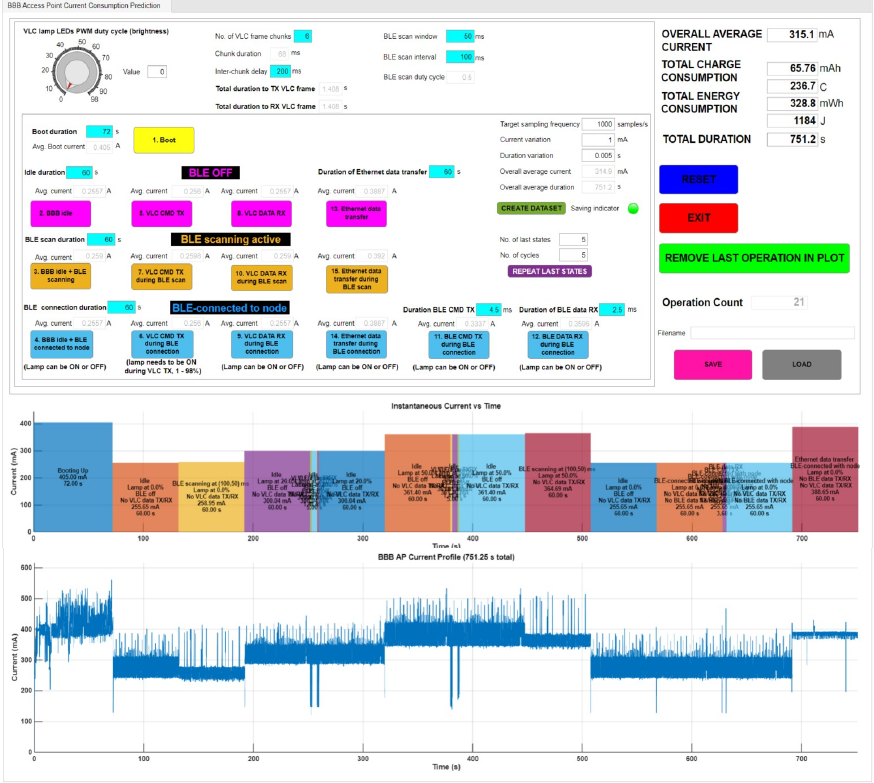}
    \caption{BBB AP energy consumption prediction App. 
    }
    \label{fig:bbbapp}
\end{figure}

\begin{figure}
    \centering    \includegraphics[width=0.75\linewidth]{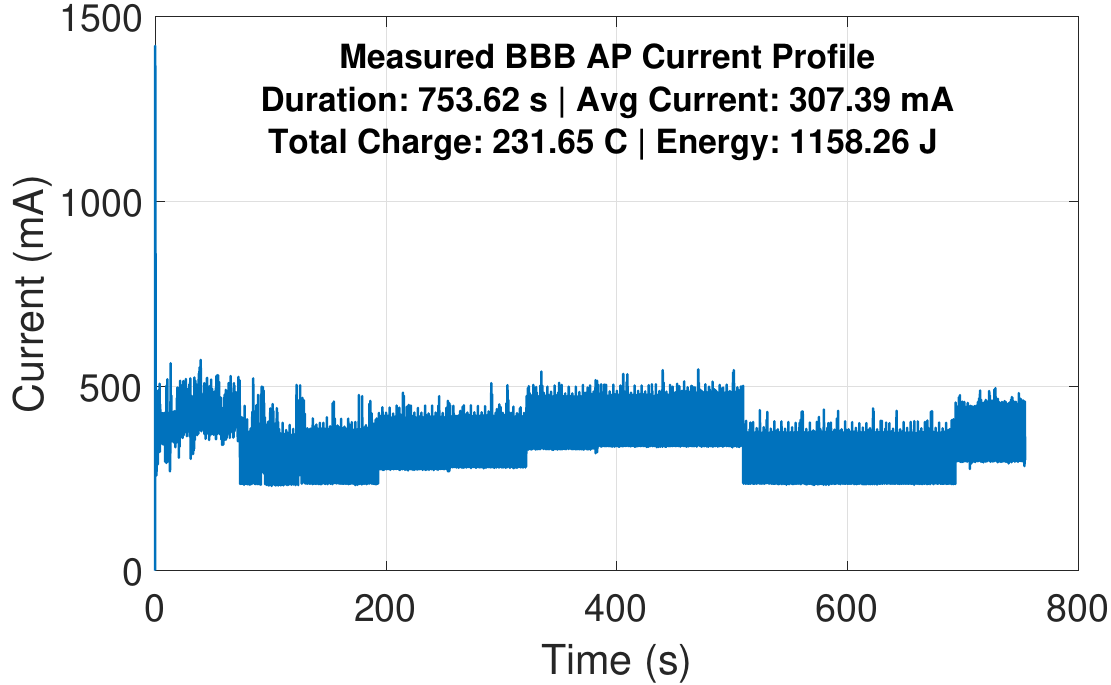}
    \caption{Measured BBB AP current profile covering a wide range of operating states. 
    }
    \label{fig:bbb_current}
\end{figure}

Similar to the node-level energy prediction tool, we developed a tool for the BBB AP, based on its energy models, as illustrated in Figure~\ref{fig:bbbapp}.
 To validate the energy prediction of the tool, 
we defined a comprehensive experimental scenario and collected real energy measurements on the BBB AP (see Figure~\ref{fig:bbb_current}) covering a wide range of operating states—from boot and idle conditions, through various VLC and BLE activity modes, to Ethernet data transmission. The evaluated states include configurations with the VLC lamp off or active at different duty cycles (e.g. 20\% and 50\%), periods of VLC command transmission and data reception, BLE scanning and connected modes with and without data exchange, and combined VLC–BLE operations.
A comparison between the measured current profile in Figure~\ref{fig:bbb_current} and the predicted results in Figure~\ref{fig:bbbapp} for the same scenario demonstrates a high level of accuracy, with all evaluated metrics achieving agreement within the 97.5–97.8\% range. These results confirm that the proposed prediction model reliably estimates the current consumption, charge, and energy of the BBB AP, closely matching the experimental measurements.

\subsection{Network-Level Energy Measurement, Analysis, Modeling and Prediction}
We designed and deployed a comprehensive measurement setup to accurately capture the actual energy consumption of the RIoT network under diverse operating conditions. Recognizing that energy demand in IoT networks depends on multiple factors, spanning node configurations and communication modalities, our setup was tailored to systematically analyze their inter-dependencies.

\subsubsection{Measurement Setup}

To address the challenge of acquiring precise, time-synchronized energy data from distributed IoT nodes and gateway/AP, we developed an automated Energy Consumption Collection System, as illustrated in Figure~\ref{fig:Energy_Consumption_Collection_System}. The system architecture consists of three main components:

\begin{figure}
    \centering    \includegraphics[width=1\linewidth]{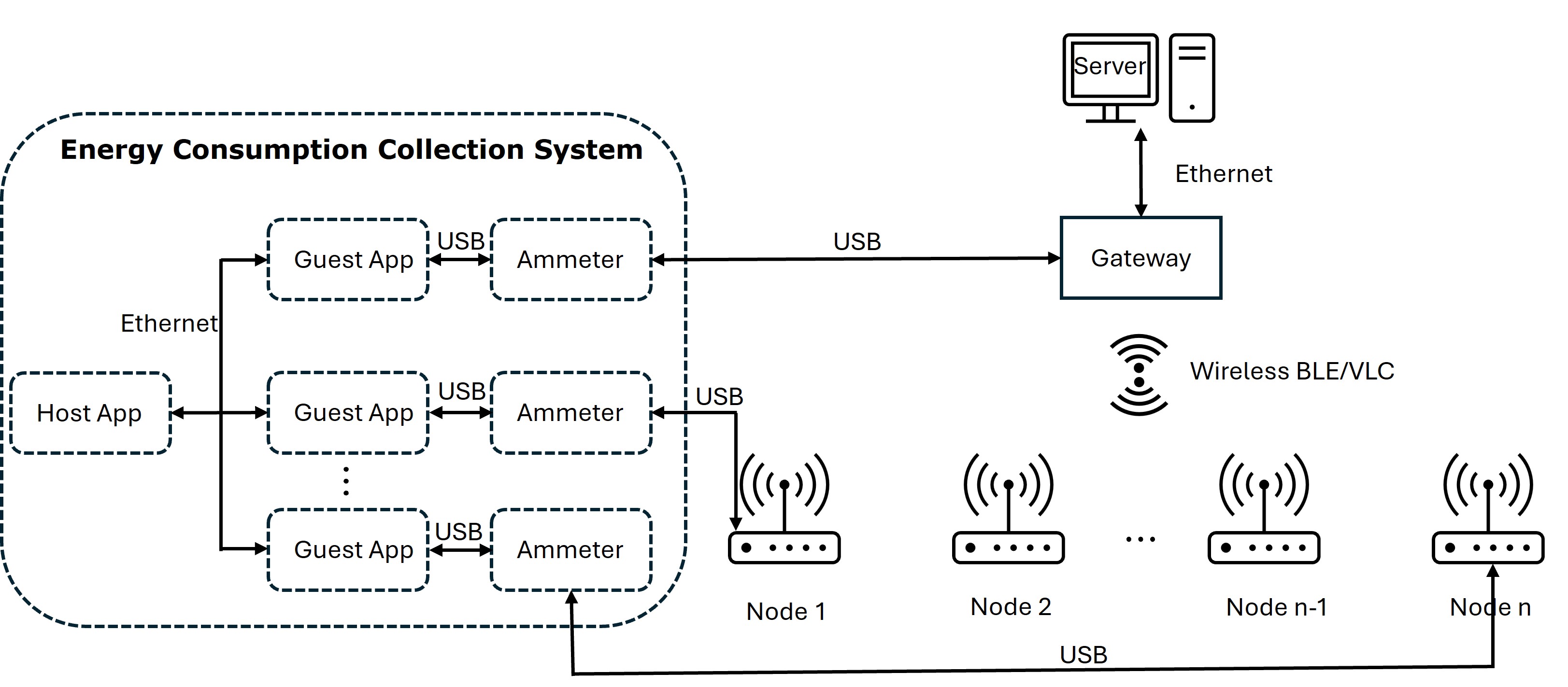}
    \caption{Energy consumption collection system. 
    }
    \label{fig:Energy_Consumption_Collection_System}
\end{figure}

\begin{itemize}
    \item Host Application: Centrally aggregates and manages data, visualizes energy consumption, and stores synchronized datasets.
    \item Guest Applications: Deployed for each IoT node and gateway to collect local current readings and relevant node-level factors (e.g., BLE/VLC configuration parameters), which are then transmitted to the Host Application.
    \item Ammeter (Power Profiler Kit II): Integrated with each node and gateway to provide current measurements, ensuring precise energy profiling.
\end{itemize}

 \begin{figure}[t]
    \centering
    \subfloat[Host Application Interface]{
        \includegraphics[width=1\linewidth]{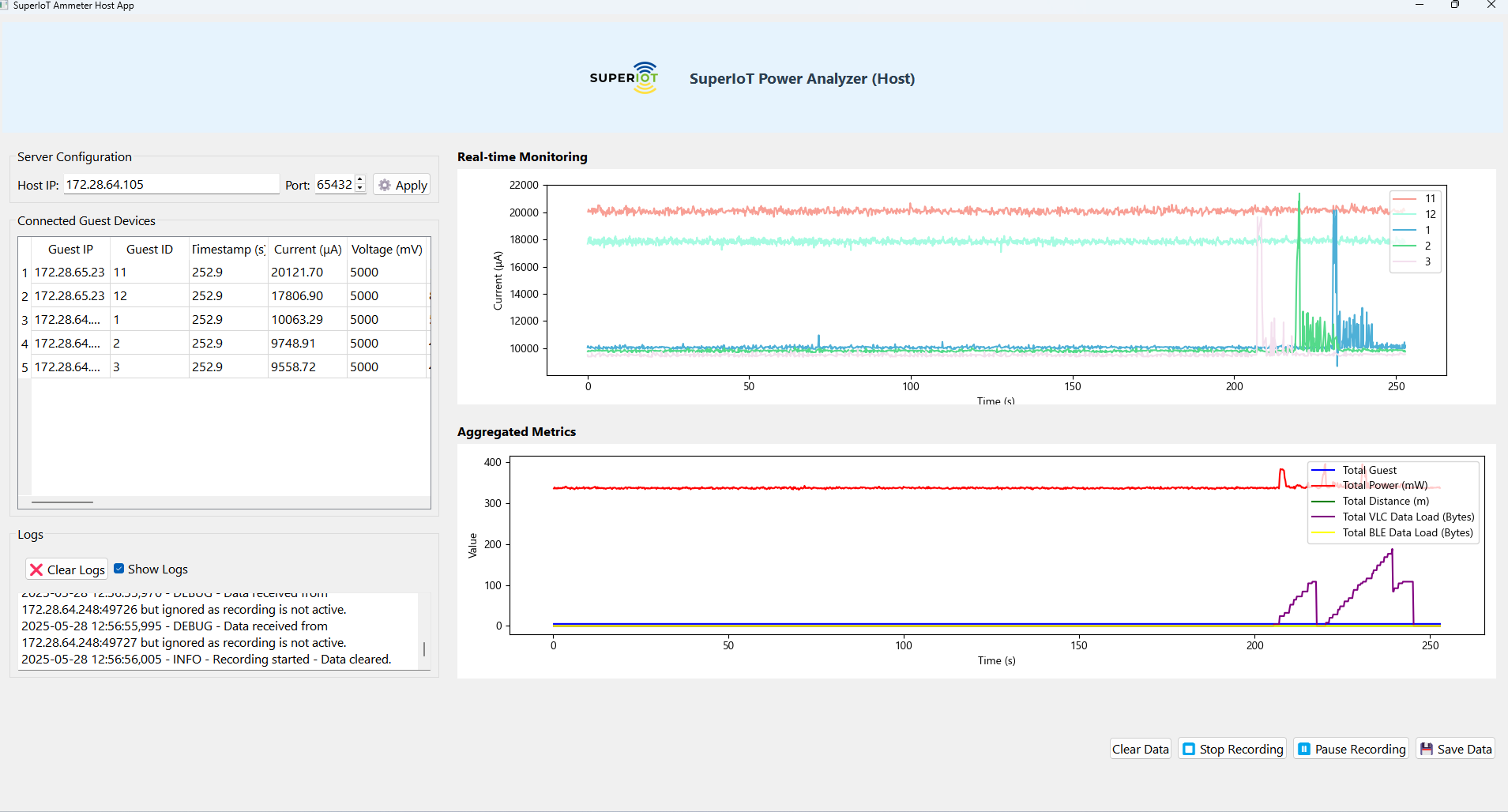}
        \label{fig:host_app}
    }\\[0.5cm]
    \subfloat[Guest Application Interface]{
        \includegraphics[width=1\linewidth]{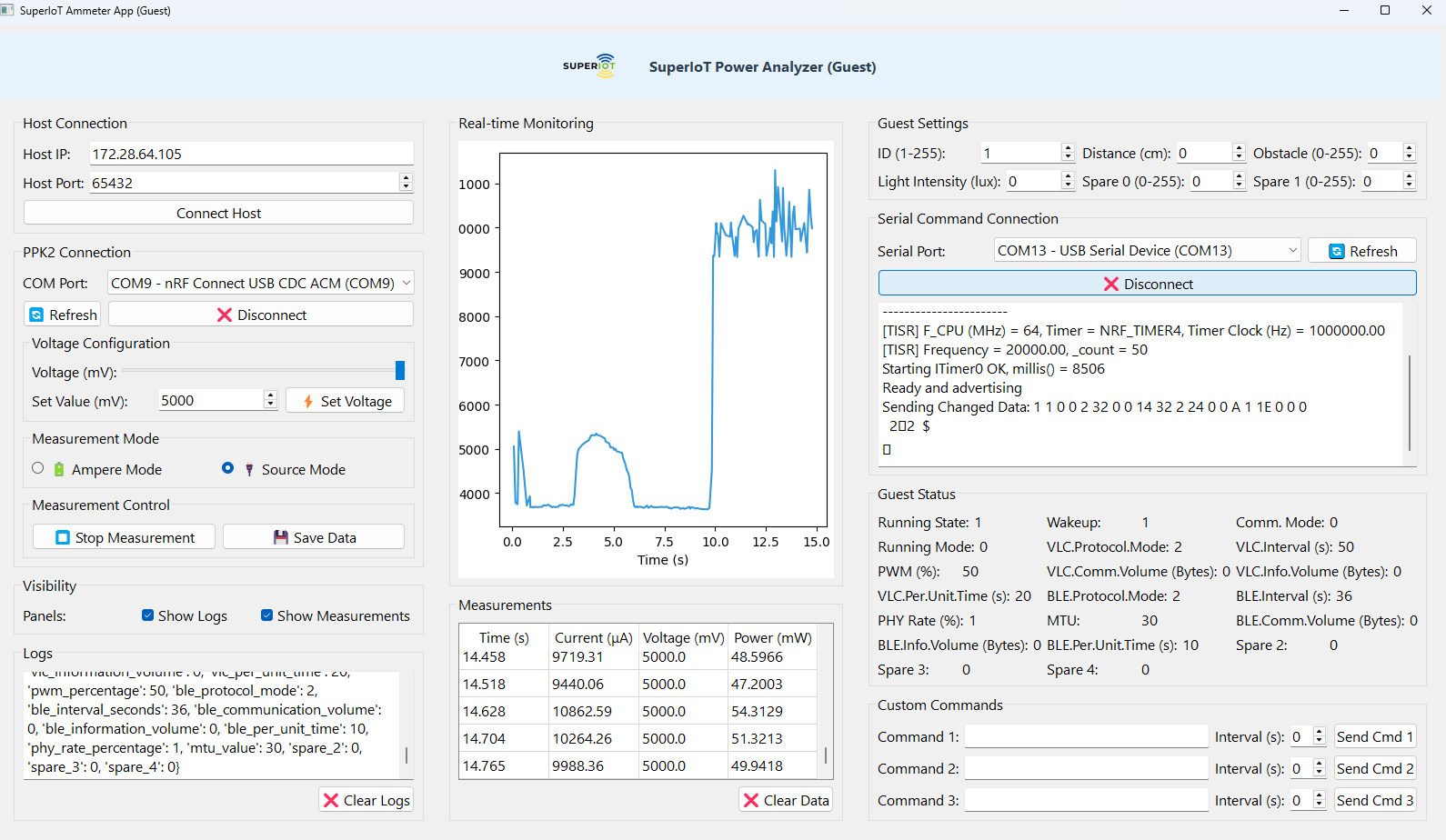}
        \label{fig:guest_app}
    }
    \caption{User interfaces for the Host and Guest Applications.}
    \label{fig:system_ui}
\end{figure}

Figure~\ref{fig:system_ui} shows the user interfaces of the Host and Guest Applications, both of which support real-time monitoring and system configuration. This modular, automated architecture enables scalable data acquisition, which is crucial for characterizing network-level energy consumption across a wide range of deployment scenarios.

\subsubsection{Dataset Preprocessing}

Based on the network-level energy measurement setup, raw data were collected from distributed nodes, including timestamped current measurements and node-specific configurations. The current dataset contains measurements from multiple RIoT nodes operating in distinct states such as idle, BLE advertising/connected modes, BLE and VLC data transmission, deep sleep, etc.
Before model training and analysis, the collected data underwent preprocessing, including synchronization of measurements across all nodes to a common time base to ensure dataset consistency, and noise filtering to remove transient spikes and measurement artifacts, thereby improving signal quality. 

\subsubsection{Machine Learning Models for Predicting Network-Level Energy Consumption}

Leveraging the preprocessed dataset, we trained multiple machine learning (ML) models to predict the RIoT network’s energy consumption under varying configurations. The prediction pipeline consists of two main components. First, the input features include the number of nodes, payload size, and the duration of operational states. 
Second, the model selection process evaluates several algorithms, including Deep Neural Network (DNN), 
Random Forest,
Gradient Boosting,
Extra Trees, 
Linear Regression, 
and Ridge Regression, 
all of which were selected for their ability to capture both linear and non-linear relationships among diverse influencing factors. 
The DNN model employed a Multi-Layer Perceptron (MLP) architecture with two hidden layers consisting of 50 and 25 neurons, respectively. It used ReLU activation functions and the L-BFGS 
optimizer with a maximum of 2000 iterations. The random forest and extra trees models were ensemble-based, each comprising 100~decision trees with bootstrap aggregation and a maximum tree depth of five. The gradient boosting model implemented sequential boosting with 100 stages, a learning rate of 0.1, and a maximum tree depth of three. For comparison, the linear regression model applied Ordinary Least Squares (OLS) with standardized input features, while the Ridge regression model introduced L2 regularization with an alpha parameter of 1.0. All models were trained to predict current consumption (µA) using three input variables: the duration of operational states, VLC payload size, and BLE payload size. Model performance was evaluated using the R\textsuperscript{2},
Mean Absolute Error (MAE),
and Root Mean Square Error (RMSE)
metrics.
\begin{table}[t]
    \centering
    \caption{Performance of different ML models for predicting RIoT network energy consumption}
    \label{tab:ml_performance}
    \footnotesize
    \begin{tabular}{lccc}
        \hline
        \textbf{Model} & \textbf{R\textsuperscript{2}} & \textbf{MAE} & \textbf{RMSE} \\
        \hline
        Ridge Regression  & 0.3084 & 2635.6 & 3111.1 \\
        Linear Regression  & 0.3119 & 2646.7 & 3103.3 \\
        Random Forest  & 0.9687 & 565.5 & 662.3 \\
        Extra Trees  & 0.9936 & 191.9 & 299.8 \\
        Gradient Boosting  & 0.9937 & 181.2 & 297.1 \\
        Neural Network & 0.9937 & 179.8 & 297.1 \\
        \hline
    \end{tabular}
\end{table}
These metrics collectively provide a comprehensive assessment of each model’s predictive accuracy and error distribution.

Table~\ref{tab:ml_performance} summarizes the performance of the ML models.
The results demonstrate that non-linear models, specifically Random Forest, Extra Trees, Gradient Boosting, and DNN, achieved excellent predictive performance, with R\textsuperscript{2} values exceeding 0.96 in most cases and relatively low MAE and RMSE. Among them, the neural network and gradient boosting models achieved the best overall results (R\textsuperscript{2}= 0.9937, MAE=179.8, RMSE=297.1). In contrast, linear models (linear regression and Ridge regression) performed significantly worse, with R\textsuperscript{2} 0.3119 and much higher errors. This highlights the complex, non-linear nature of the relationship between network energy consumption and factors like the number of nodes, payload size, and BLE/VLC configurations. These findings confirm that advanced ML models can accurately predict network-level energy consumption across diverse configurations.

\section{Conclusion}
This work extends our ongoing research on energy profiling in reconfigurable IoT systems by advancing from node-level characterization to a unified, network-level framework for empirical measurement and prediction. In this extended study, we integrate custom-engineered gateway and access point platforms, providing a holistic analysis of energy consumption across the entire IoT ecosystem.
The proposed framework enables synchronized, high-fidelity energy data acquisition from multiple nodes and gateways under diverse operating conditions, capturing coordinated interactions across heterogeneous subsystems. It further includes advanced estimation and prediction tools that generate time-aligned, machine learning–ready datasets for AI-driven energy forecasting and adaptive optimization. Validation within a real-world IoT testbed demonstrates the framework’s robustness and scalability, highlighting its potential to inform the design of next-generation, energy-efficient IoT architectures.

\section{Acknowledgments}
The SUPERIOT project has received funding from the Smart Networks and Services Joint Undertaking (SNS JU) under the European Union’s Horizon Europe research and innovation programme under Grant Agreement No 101096021, including funding under the UK government’s Horizon Europe funding guarantee, UKRI Grant Reference Number 10053751. 

Views and opinions expressed are however those of the authors only and do not necessarily reflect those of the European Union, SNS JU or UKRI. The European Union, SNS JU or UKRI cannot be held responsible for them.

\printbibliography[title={References}]

\end{document}